\documentclass[aps,showpacs,showkeys]{revtex4}
\usepackage{amsmath,amsthm,amssymb,epsfig,alltt}

\begin{document}

\def\a{\alpha}
\def\b{\beta}
\def\d{{\delta}}
\def\l{\lambda}
\def\e{\epsilon}
\def\p{\partial}
\def\m{\mu}
\def\n{\nu}
\def\t{\tau}
\def\th{\theta}
\def\s{\sigma}
\def\g{\gamma}
\def\o{\omega}
\def\r{\rho}
\def\z{\zeta}
\def\D{\Delta}
\def\half{\frac{1}{2}}
\def\hatt{{\hat t}}
\def\hatx{{\hat x}}
\def\hatp{{\hat p}}
\def\hatX{{\hat X}}
\def\hatY{{\hat Y}}
\def\hatP{{\hat P}}
\def\haty{{\hat y}}
\def\whatX{{\widehat{X}}}
\def\whata{{\widehat{\alpha}}}
\def\whatb{{\widehat{\beta}}}
\def\whatV{{\widehat{V}}}
\def\hatth{{\hat \theta}}
\def\hatta{{\hat \tau}}
\def\hatrh{{\hat \rho}}
\def\hatva{{\hat \varphi}}
\def\barx{{\bar x}}
\def\bary{{\bar y}}
\def\barz{{\bar z}}
\def\baro{{\bar \omega}}
\def\barpsi{{\bar \psi}}
\def\sp{\sigma^\prime}
\def\nn{\nonumber}
\def\cb{{\cal B}}
\def\2pap{2\pi\alpha^\prime}
\def\wideA{\widehat{A}}
\def\wideF{\widehat{F}}
\def\beq{\begin{eqnarray}}
 \def\eeq{\end{eqnarray}}
 \def\4pap{4\pi\a^\prime}
 \def\op{\omega^\prime}
 \def\xp{{x^\prime}}
 \def\sp{{\s^\prime}}
 \def\ap{{\a^\prime}}
 \def\tp{{\t^\prime}}
 \def\zp{{z^\prime}}
 \def\xpp{x^{\prime\prime}}
 \def\xppp{x^{\prime\prime\prime}}
 \def\barxp{{\bar x}^\prime}
 \def\barxpp{{\bar x}^{\prime\prime}}
 \def\barxppp{{\bar x}^{\prime\prime\prime}}
 \def\zetap{{\zeta^\prime}}
 \def\barchi{{\bar \chi}}
 \def\baro{{\bar \omega}}
 \def\bpsi{{\bar \psi}}
 \def\barg{{\bar g}}
 \def\barz{{\bar z}}
 \def\bareta{{\bar \eta}}
 \def\ta{{\tilde \a}}
 \def\tb{{\tilde \b}}
 \def\tc{{\tilde c}}
 \def\tz{{\tilde z}}
 \def\tJ{{\tilde J}}
 \def\tpsi{\tilde{\psi}}
 \def\tal{{\tilde \alpha}}
 \def\tbe{{\tilde \beta}}
 \def\tga{{\tilde \gamma}}
 \def\tchi{{\tilde{\chi}}}
 \def\tx{{\tilde x}}
 \def\barth{{\bar \theta}}
 \def\bareta{{\bar \eta}}
 \def\barom{{\bar \omega}}
 \def\bole{{\boldsymbol \epsilon}}
 \def\bolth{{\boldsymbol \theta}}
 \def\bomega{{\boldsymbol \omega}}
 \def\bolmu{{\boldsymbol \mu}}
 \def\bolal{{\boldsymbol \alpha}}
 \def\bolbe{{\boldsymbol \beta}}
 \def\bolL{{\boldsymbol  L}}
 \def\bolX{{\boldsymbol X}}
 \def\boln{{\boldsymbol n}}
 \def\bols{{\boldsymbol s}}
 \def\bolS{{\boldsymbol S}}
 \def\bola{{\boldsymbol a}}
 \def\bolA{{\boldsymbol A}}
 \def\bolJ{{\boldsymbol J}}
 \def\tr{{\rm tr}}
 \def\bbp{{\boldsymbol p}}
 \def\bbphi{{\boldsymbol \phi}}
 \def\bbA{{\boldsymbol A}}
 \def\mathP{{\mathbb P}}
 \def\mathN{{\boldsymbol N}}
 \def\mathN{{\mathbb N}}
 \def\bbP{{\boldsymbol P}}
 \def\bbh{{\boldsymbol h}}

\setcounter{page}{1}
\title[]{Graviton and Massive Symmetric Rank-Two Tensor in String Theory
}

\author{Taejin Lee\footnote{Corresponding author}}
\email{taejin@kangwon.ac.kr}
\affiliation{
Department of Physics, Kangwon National University, Chuncheon 24341
Korea}
\author{Hayun Park}
\email{nagato10@naver.com}
\affiliation{
Department of Physics, Kangwon National University, Chuncheon 24341
Korea}


\begin{abstract}
Spin-two particles appear in the spectra of both open and closed string theories. We studied a graviton and massive symmetric rank-two tensor in string theory, both of which carry spin two.
A graviton is a massless spin-two particle in closed string theory while a symmetric rank-two tensor 
is a massive particle with spin two in open string theory. 
Using Polyakov's string path integral formulation of string scattering amplitudes, 
we calculated cubic interactions of both spin-two particles explicitly, including $\ap$-corrections (string corrections).
We observed that the cubic interactions of the massive spin-two particle differed from those of the graviton. The massive symmetric rank-two tensor in open string theory becomes massless in the high energy limit where $\ap \rightarrow \infty$
and $\ap$-correction terms, containing  higher derivatives, dominate: In this limit the local cubic action of the symmetric rank-two tensor of open string theory coincides with that of the graviton in closed string theory.
\end{abstract}




\maketitle

\section{Introduction}

It is interesting to note that spin-two particles appear in the spectra of both open and closed string theories. The differences between these
two particles lie in their masses. The spin-two particle contained in the spectrum of closed string is massless, while that belonging to the spectrum of open string theory is massive. To distinguish one from the other, we refer to 
the former as a graviton and the latter as a symmetric rank-two tensor throughout this work. Because it is presumed that string theories are consistent and finite, we considered them as guiding principles to study the cubic interactions of both these spin-two particles. 
The field theory for massive spin-two particles was developed long ago in 1939 by Fierz and Pauli \cite{FierzP1939}, who introduced a particular form of the mass term for the spin-two field in the linearized Einstein gravity. With this particular form of the mass term, 
the Fierz--Pauli (FP) model is free of ghost, named the Boulware--Deser ghost \cite{BoulwareD1972,BoulwareD1975PL,BoulwareD1975NP} 
having a negative kinetic energy; however, the model does not reduce to the linearized Einstein 
gravity in the massless limit. The massless limit of the FP model is singular, and a discontinuity, known as the van Dam--Veltman--Zakharov (vDVZ) discontinuity \cite{vanDam1970,Zakharov1970}, is encountered. Moreover, if we introduce interaction terms to the FP model, the Boulware--Deser ghost  begins to reappear. These problems in the FP model may be resolved through Vainstein screening \cite{Vainshtein1972}. 

In a recent study \cite{ParkLee2019a}, we proposed open string theory to study massive gravity and performed canonical 
quantization of the symmetric rank-two tensor model of Siegel and Zwiebach (SZ model) containing two Stueckelberg fields. 
The advantage of using the SZ model to describe massive spin-two particles is that this model has local gauge symmetry,
through which the FP model may be transformed into various other gauge equivalent forms. 
When choosing a particular gauge to be imposed on the Stueckelberg fields, the SZ model reduces to the FP model. 
As the SZ model is a gauge invariant theory, we may choose various alternative gauge conditions. The most convenient 
gauge condition may be the transverse-traceless (TT) gauge, where the two Stueckelberg fields can be completely decoupled from 
the symmetric rank-two tensor to be integrated trivially. In the TT gauge, the SZ model, describing the massive spin-two field, 
smoothly reduces to the linearized Einstein gravity without encountering any discontinuity.  

The study presented in this work is the extension of our previous study on the free field model of massive symmetric rank-two tensors to 
interacting models for calculating cubic couplings of the rank-two tensor field explicitly within the framework of string theory. 
For comparison, we evaluated the cubic couplings of the graviton in closed string theory and those of the massive 
symmetric rank-two tensor in open string theory. The main method for evaluating the cubic interactions was Polyakov's string path integral formula of scattering amplitudes,
\cite{Polyakov1981},
which is discussed in detail elsewhere \cite{TLeeJKPS2017,Lee2017d,TLee2017cov,SHLai2018,TLee2019PL}. 
In the proper-time gauge \cite{TLeeann1988}, we cast the three-string scattering amplitudes
into the Feynman--Schwinger proper-time representation of quantum field theory, from which the cubic interactions of the spin-two fields 
could be directly obtained. The cubic graviton interaction terms, including $\ap$-corrections, were calculated using the Polyakov integral formulation of three closed-string scattering amplitudes. Then, we evaluated the cubic couplings of massive symmetric rank-two tensors, applying the same procedure using Polyakov's string path integral formulation of three open-string scattering amplitudes. It was observed that the cubic interactions of the symmetric rank-two tensors differed from those of gravitons. However, in the massless limit where $\ap \rightarrow \infty$, $\ap$-correction terms dominate and the local cubic action of the symmetric rank-two tensor field in open string theory becomes identical to that of the graviton in closed string theory.

\section{Cubic Interactions of Graviton in Closed String Theory}

We may express the three-string scattering amplitudes using Polyakov's string path integral formulation of the
string worldsheet theory with three external closed strings as follows:
\begin{subequations}
\beq
{\cal A}_{[3]} &=& \frac{2g_{\rm closed}}{3} \int D[X]D[h] \exp \left(iS + i \int_{\p M} \sum_{r=1}^3  P^{(r)}\cdot X^{(r)} d\s \right), \label{W4}\\
S &=& -\frac{1}{4\pi} \int_M d\t d\s \sqrt{-h} h^{\a\b} \frac{\p X^\m}{\p \s^\a} \frac{\p X^\n}{\p \s^\b} \eta_{\m\n}, ~~~~~ \m, \n = 0, \dots , d-1  
\eeq
\end{subequations}
where 
$\s^1= \t$, $\s^2 = \s$. Here, $d = 26$ for the bosonic string and $d = 10$ for the super-string. 
If we evaluate the three closed-string scattering amplitudes in the proper-time gauge and cast them into the Feynman--Schwinger proper-time
representation, we obtain \cite{TLeeEPJ2018}
\begin{subequations}
\beq \label{3string}
{\cal A}_{[3]C} &=&  \frac{2g_{\rm closed}}{3} \int \prod_{r=1}^3 dp^{(r)} \d \left(\sum_{r=1}^3 p^{(r)} \right) \langle \Psi_1, \Psi_2, \Psi_3 \vert E_{[3]\text{closed}} \vert 0 \rangle, \\
E_{[3]\text{closed}}\vert 0 \rangle
&=& \exp  \Biggl\{
\sum_{r,s=1}^3 \Bigl( \sum_{n, m \ge 1} \frac{1}{2} \bar N^{rs}_{nm}\,\frac{\a^{(r)\dag}_n}{2} \cdot 
\frac{\a^{(r)\dag}_m}{2} 
+ \sum_{n \ge 1}\bar N^{rs}_{n0} \frac{\a^{(r)\dag}_n}{2} \cdot p^{(s)}\Bigr) 
\Biggr\} \nn\\
&&  \exp \Biggl\{ \t_0 \sum_{r=1}^3 \frac{1}{\a_r} \left(\half \left(p^{(r)} \right)^2 -1 \right) 
\Biggr\}  \nn\\
&& \exp  \Biggl\{
\sum_{r,s=1}^3 \Bigl( \sum_{n, m \ge 1} \frac{1}{2} \bar N^{rs}_{nm}\,\frac{\tilde\a^{(r)\dag}_n}{2} \cdot \frac{\tilde\a^{(r)\dag}_m}{2} 
+ \sum_{n \ge 1}\bar N^{rs}_{n0} \frac{\tilde\a^{(r)\dag}_n}{2} \cdot p^{(s)} \Bigr) \Biggr\}\nn\\
&& \exp \Biggl\{ \t_0 \sum_{r=1}^3 \frac{1}{\a_r} \left(\half \left(p^{(r)}\right)^2 -1 \right) 
\Biggr\} \vert 0 \rangle, \label{3vertex}
\eeq  
\end{subequations}
where $p= p^{(2)}- p^{(1)} $, $\t_0 = -2\ln 2$, and  $\a_1=\a_2=1, ~\a_3= -2$ in the proper-time gauge. 
Here, $\bar N^{rs}_{nm}$ and $\bar N^r_n$ are the Neumann functions of the three open-string scattering amplitudes \cite{TLeeJKPS2017}. 

\begin{figure}[htbp]
   \begin {center}
    \epsfxsize=0.3\hsize

	\epsfbox{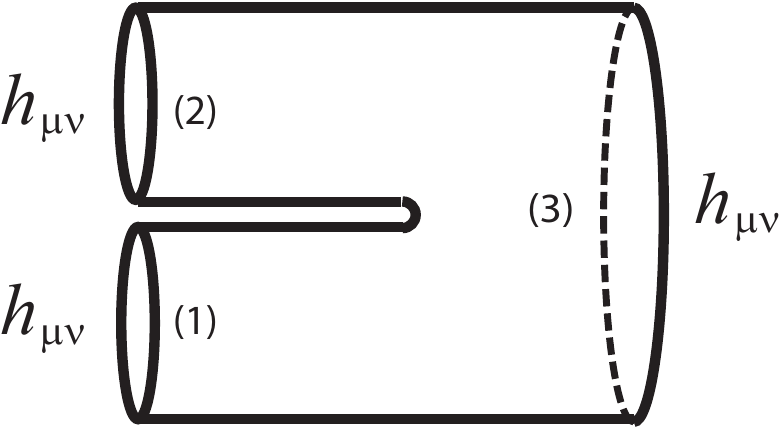}
   \end {center}
   \caption{\label{closedh3} Three-graviton interactions in the proper-time gauge.}
\end{figure}

It is noteworthy that the
Three closed-string amplitudes may be completely factorized into three open-string scattering amplitudes to confirm the Kawai--Lewellen--Tye
(KLT) relation \cite{Kawai86} at the level of second quantized string theory. 
Choosing the external three closed-string states as 
$\langle \Psi^{(1)}, \Psi^{(2)}, \Psi^{(3)} \vert = \langle 0 \vert \prod_{r=1}^3 
\left(h_{\m\n}(p^{(r)}) a^{(r)\m}_{1} \tilde  a^{(r)\n}_{1} \right)$, we obtained the three-graviton scattering amplitudes, depicted in
Fig. \ref{closedh3} 
\beq
{\cal A}_{hhh,C} &=& {\cal A}_{hhh,C}^{(1,1)} + {\cal A}_{hhh,C}^{(1,3)} + {\cal A}_{hhh,C}^{(3,1)} + {\cal A}_{hhh,C}^{(3,3)}.
\eeq
Here, with $\bbh(r) =h_{\m\n}(p^{(r)}) a^{(r)\m}_{1} \tilde  a^{(r)\n}_{1}$, $a_1^\dag = \a_1^\dag$,  
\begin{subequations}
\beq
{\cal A}_{hhh,C}^{(1,1)} &=& \frac{2^7g_{\rm closed}}{3}\int \prod_{r=1}^3 dp^{(r)} \d \left(\sum_{r=1}^3 p^{(r)}\right) \,
\langle 0 \vert \bbh(1)\bbh(2) \bbh(3) \nn\\
&&\left\{\half \sum_{r,s=1}^3 \bar N^{rs}_{11} \frac{a^{(r)\dagger}_{1}}{2} \cdot \frac{a^{(s)\dagger}_{1}}{2}
\left(\sum_{r=1}^3\bar N^r_1 \frac{a^{(r)\dagger}_{1}}{2} \cdot p \right)\right\}
\left\{\half \sum_{r,s=1}^3 \bar N^{rs}_{11} \frac{\tilde a^{(r)\dagger}_{1}}{2} \cdot \frac{\tilde a^{(s)\dagger}_{1}}{2}
\left(\sum_{r=1}^3\bar N^r_1 \frac{\tilde a^{(r)\dagger}_{1}}{2} \cdot p \right) \right\} 
\vert 0 \rangle, \\
{\cal A}_{hhh,C}^{(1,3)} &=&  \frac{2^7g_{\rm closed}}{3}\int \prod_{r=1}^3 dp^{(r)} \d \left(\sum_{r=1}^3 p^{(r)}\right) \,
\langle 0 \vert \bbh(1)\bbh(2) \bbh(3) \nn\\
&&\left\{\half \sum_{r,s=1}^3 \bar N^{rs}_{11} \frac{a^{(r)\dagger}_{1}}{2} \cdot \frac{a^{(s)\dagger}_{1}}{2}
\left(\sum_{r=1}^3\bar N^r_1 \frac{a^{(r)\dagger}_{1}}{2} \cdot p \right)\right\} 
\left\{\frac{1}{3!}\left(\sum_{r=1}^3\bar N^r_1 \frac{\tilde a^{(r)\dagger}_{1}}{2} \cdot p\right)^3 
\right\} \vert 0 \rangle, \\
{\cal A}_{hhh,C}^{(3,1)} &=& \frac{2^7g_{\rm closed}}{3}\int \prod_{r=1}^3 dp^{(r)} \d \left(\sum_{r=1}^3 p^{(r)}\right) \,
\langle 0 \vert \bbh(1)\bbh(2) \bbh(3) \nn\\
&&\left\{ \frac{1}{3!}\left(\sum_{r=1}^3\bar N^r_1 \frac{a^{(r)\dagger}_{1}}{2} \cdot p \right)^3 \right\} \left\{\half \sum_{r,s=1}^3 \bar N^{rs}_{11} \frac{\tilde a^{(r)\dagger}_{1}}{2} \cdot \frac{\tilde a^{(s)\dagger}_{1}}{2}
\left(\sum_{r=1}^3\bar N^r_1 \frac{\tilde a^{(r)\dagger}_{1}}{2} \cdot p \right) \right\} 
\vert 0 \rangle, \\
{\cal A}_{hhh,C}^{(3,3)} &=& \frac{2^7g_{\rm closed}}{3}\int \prod_{r=1}^3 dp^{(r)} \d \left(\sum_{r=1}^3 p^{(r)}\right) \,
\langle 0 \vert \bbh(1)\bbh(2) \bbh(3) \nn\\
&&\left\{ \frac{1}{3!}\left(\sum_{r=1}^3\bar N^r_1 \frac{a^{(r)\dagger}_{1}}{2} \cdot p \right)^3\right\} \left\{\frac{1}{3!}\left(\sum_{r=1}^3\bar N^r_1 \frac{\tilde a^{(r)\dagger}_{1}}{2} \cdot p \right)^3 
\right\} \vert 0 \rangle .
\eeq 
\end{subequations}
We observed that ${\cal A}_{hhh}^{(1,3)}$ and ${\cal A}_{hhh}^{(3,1)}$ correspond to $\ap$-corrections of order ${\cal O}(\ap)$ and 
${\cal A}_{hhh,C}^{(3,3)}$ to an $\ap$-correction of order ${\cal O}(\ap^2)$ in the three-graviton scattering amplitude 
of Einstein's gravity ${\cal A}_{hhh,C}^{(1,1)}$.  Through algebraic calculations, we obtained
\beq
{\cal A}_{hhh,C}^{(1,1)} 
&=& \kappa
\int \prod_{i=1}^3 dp^{(r)} \d\left(\sum_{r=1}^3 p^{(r)} \right)  h_{\m_{1}\n_{1}}(p^{(1)})h_{\m_{2}\n_{2}}(p^{(2)})h_{\m_{3}\n_{3}}(p^{(3)}) \nn\\
&& \Bigl\{
\eta^{\m_1\m_2} p^{(1)\m_3} + \eta^{\m_2\m_3} p^{(2)\m_1} 
+ \eta^{\m_3\m_1} p^{(3)\m_2} \Bigr\}\Bigl\{\eta^{\n_1\n_2} p^{(1)\n_3} + \eta^{\n_2\n_3} p^{(2)\n_1} 
+ \eta^{\n_3\n_1} p^{(3)\n_2} \Bigr\}, 
\eeq  
where $\kappa = \frac{g_{\rm closed}}{3\cdot 2^5} $. This is precisely the three-graviton interaction term of the conventional perturbative Einstein's gravity \cite{Dewitt1,Dewitt2,Dewitt3,Scherk74,Sannan86}.
Similar algebraic calculations yielded the following cubic interaction of gravitons in string theory, including $\ap$-corrections:
\beq
{\cal A}_{hhh,C}^{(1,1)} + {\cal A}_{hhh,C}^{(1,3)} + {\cal A}_{hhh,C}^{(3,1)} + {\cal A}_{hhh,C}^{(3,3)} 
&=& \kappa\int \prod_{i=1}^3 dp^{(i)} \d\left(\sum_{i=1}^3 p^{(i)} \right)  h_{\m_{1}\n_{1}}(p^{(1)})h_{\m_{2}\n_{2}}(p^{(2)})h_{\m_{3}\n_{3}}(p^{(3)}) \nn\\
&&
\Bigl(\eta^{\m_1\m_2} p^{(1)\m_3} + \eta^{\m_2\m_3} p^{(2)\m_1} 
+ \eta^{\m_3\m_1} p^{(3)\m_2} + p^{(1)\m_3} p^{(2)\m_1} p^{(3)\m_2} \Bigr) \nn\\
&& \Bigl(\eta^{\n_1\n_2} p^{(1)\n_3} + \eta^{\n_2\n_3} p^{(2)\n_1} 
+ \eta^{\n_3\n_1} p^{(3)\n_2} + p^{(1)\n_3} p^{(2)\n_1} p^{(3)\n_2}\Bigr).
\eeq
This result, exhibiting clearly the KLT relation between cubic couplings of open and closed string theories, 
is in perfect agreement with previous results \cite{GreenSW1987}.

\section{Cubic Interactions of Massive Symmetric Rank-Two Tensor}

\begin{figure}[htbp]
   \begin {center}
    \epsfxsize=0.3\hsize

	\epsfbox{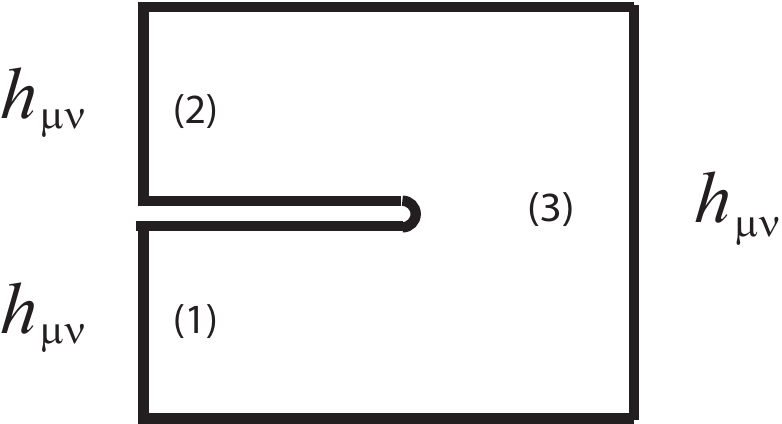}
   \end {center}
   \caption{\label{openh3} Three rank-two tensor interactions in the proper-time gauge.}
\end{figure}

The three open-string scattering amplitudes, if cast into the Feynman--Schwinger proper-time representation, may be written as follows
\cite{TLeeJKPS2017}: 
\begin{subequations}
\beq \label{open3}
{\cal A}_{[3],O} &=&  \frac{2g_{\rm open}}{3} \int \prod_{r=1}^3 dp^{(r)} \d \left(\sum_{r=1}^3 p^{(r)} \right) 
\langle \Psi_1, \Psi_2, \Psi_3 \vert E_{[3]\text{open}}  \vert 0 \rangle, \label{3string1} \\
E_{[3]\text{open}} 
&=& \exp \,\Biggl\{ \frac{1}{2} \sum_{r,s =1}^3 \sum_{n, m \ge 1} \bar N^{rs}_{nm} \, 
\a^{(r)\dagger}_{n} \cdot \a^{(s)\dagger}_{m}  + 
\sum_{r=1}^3 \sum_{n \ge 1} \bar N^r_n \a^{(r)\dag}_{n} \cdot p +\t_0 \sum_{r=1}^3 \frac{1}{\a_r} \left(\frac{\left(p^{(r)}\right)^2}{2} -1 \right)
\Biggr\} , \label{3string2}
\eeq
\end{subequations}
where $\t_0 =-2\ln 2$ and $p = p^{(2)}- p^{(1)}$. We selected the external string state, representing three rank-two tensors,
$\langle \Psi^{(1)}, \Psi^{(2)}, \Psi^{(3)} \vert = \langle 0 \vert \prod_{r=1}^3 
\left(h_{\m\n}(p^{(r)}) a^{(r)\m}_{1} \a^{(r)\n}_{1} \right)$ to calculate the three massive-symmetric-tensor scattering amplitudes,
 depicted in Fig. \ref{openh3}. 

Here, the massive symmetric rank-two tensor field, $h_{\m\n}$, satisfied the following TT gauge condition: 
$h_{\m\n}(p)p^\m = h_{\m\n}(p)p^\n =0$, $h^\m{}_\m(p)= 0$. We decomposed the scattering amplitude of the three rank-two tensors, ${\cal A}_{hhh,O}$, into four terms by expanding the three-string vertex, $E[1,2,3]_{\rm open}$, as follows:
\beq
{\cal A}_{hhh,O} &=& {\cal A}_{hhh,O}^{(3,0)}+ 
{\cal A}_{hhh,O}^{(2,2)} + {\cal A}_{hhh,O}^{(1,4)} + {\cal A}_{hhh,O}^{(0,6)}.
\eeq
To be explicit, with $\bbh(r) =h_{\m\n}(p^{(r)}) a^{(r)\m}_{1} a^{(r)\n}_{1}$
\begin{subequations}
\beq
{\cal A}_{hhh,O}^{(3,0)} &=&  \frac{2^7g_{\rm open}}{3\cdot 3!} \int \prod_{r=1}^3 dp^{(r)} \d \left(\sum_{r=1}^3 p^{(r)} \right) \langle 0 \vert \bbh(1)\bbh(2) \bbh(3) \left(\half \sum_{r,s}\bar N^{rs}_{11}a_1^{(r)\dag}\cdot a_1^{(s)\dag} \right)^3 \vert 0 \rangle, \\
{\cal A}_{hhh,O}^{(2,2)} &=& \frac{2^7g_{\rm open}}{3\cdot 2! \cdot 2!} \int \prod_{r=1}^3 dp^{(r)} \d \left(\sum_{r=1}^3 p^{(r)} \right)\langle 0 \vert \bbh(1)\bbh(2) \bbh(3) \nn\\
&& \left(\half \sum_{r,s}\bar N^{rs}_{11}a_1^{(r)\dag}\cdot a_1^{(s)\dag} \right)^2 \left(\sum_r \bar N^r_1 a^{(r)\dag}_{1}\cdot p \right)^2\vert 0 \rangle, \\
{\cal A}_{hhh,O}^{(1,4)} &=& \frac{2^7g_{\rm open}}{3\cdot 4!} \int \prod_{r=1}^3 dp^{(r)} \d \left(\sum_{r=1}^3 p^{(r)} \right)\langle 0 \vert \bbh(1)\bbh(2) \bbh(3) \nn\\
&&\left(\half \sum_{r,s}\bar N^{rs}_{11}a_1^{(r)\dag}\cdot a_1^{(s)\dag} \right) \left(\sum_r \bar N^r_1 a^{(r)\dag}_{1}\cdot p \right)^4\vert 0 \rangle, \\
{\cal A}_{hhh,O}^{(0,6)} &=& \frac{2^7g_{\rm open}}{3\cdot 6!} \int \prod_{r=1}^3 dp^{(r)} \d \left(\sum_{r=1}^3 p^{(r)} \right) \langle 0 \vert \bbh(1)\bbh(2) \bbh(3) \left(\sum_r \bar N^r_1 a^{(r)\dag}_{1}\cdot p \right)^6\vert 0 \rangle .
\eeq 
\end{subequations}

We observed that ${\cal A}_{hhh,O}^{(3,0)}$ did not contain any derivative and could not be factorized. Therefore,
\beq
{\cal A}_{hhh,O}^{(3,0)}  = 4\kappa
\int \prod_{i=1}^3 dp^{(r)} \d\left(\sum_{r=1}^3 p^{(r)} \right)  h_{\m_{1}\n_{1}}(p^{(1)})h_{\m_{2}\n_{2}}(p^{(2)})h_{\m_{3}\n_{3}}(p^{(3)}) 
\eta^{\m_1\n_2}\eta^{\m_2\n_3}\eta^{\m_3\n_1}, \label{A30}
\eeq 
where $\kappa = \frac{2^2 g_{\rm open}}{3}$. Through algebraic calculations, we also evaluated the remaining terms as follows:
\beq
{\cal A}_{hhh,O}^{(2,2)} &=&\kappa
\int \prod_{i=1}^3 dp^{(r)} \d\left(\sum_{r=1}^3 p^{(r)} \right)  h_{\m_{1}\n_{1}}(p^{(1)})h_{\m_{2}\n_{2}}(p^{(2)})h_{\m_{3}\n_{3}}(p^{(3)})\nn\\
&& \Biggl\{\biggl(\eta^{\m_1\m_2}p^{(1)\m_3}+\eta^{\m_2\m_3}p^{(2)\m_1}+\eta^{\m_3\m_1}p^{(3)\m_2}\biggr)
\biggl(\eta^{\n_1\n_2}p^{(1)\n_3}+\eta^{\n_2\n_3}p^{(2)\n_1}+\eta^{\n_3\m_1}p^{(3)\n_2}\biggr) \nn\\
&& + 2\eta^{\m_1\m_2}p^{(1)\m_3}\eta^{\n_2\n_3}p^{(2)\n_1}+ 2\eta^{\m_2\m_3}p^{(2)\m_1}\eta^{\n_3\n_1}p^{(3)\n_2}
+2\eta^{\m_3\m_1}p^{(3)\m_2}\eta^{\n_2\n_3}p^{(1)\n_3}\Biggr\}.
\eeq 
We observe that ${\cal A}_{hhh,O}^{(2,2)}$ is similar to ${\cal A}_{hhh,C}^{(1,1)}$ of closed string but differs from it:
${\cal A}_{hhh,O}^{(2,2)}$ is not factorized in contrast to ${\cal A}_{hhh,C}^{(1,1)}$. 

 It may be interesting to evaluate other terms explicitly and 
compare them with their counterparts of closed string. Using the symmetric property, $h_{\m\n} = h_{\n\m}$, and the TT gauge 
conditions, we obtained 
\beq
{\cal A}_{hhh,O}^{(1,4)} &=&\kappa
\int \prod_{i=1}^3 dp^{(r)} \d\left(\sum_{r=1}^3 p^{(r)} \right)  h_{\m_{1}\n_{1}}(p^{(1)})h_{\m_{2}\n_{2}}(p^{(2)})h_{\m_{3}\n_{3}}(p^{(3)})\nn\\
&& \Biggl\{\biggl(\eta^{\m_1\m_2}p^{(1)\m_3}+\eta^{\m_2\m_3}p^{(2)\m_1}+\eta^{\m_3\m_1}p^{(3)\m_2}\biggr)p^{(1)\n_3}p^{(2)\n_1}p^{(3)\n_2} \nn\\
&& + \biggl(\eta^{\n_1\n_2}p^{(1)\n_3}+\eta^{\n_2\n_3}p^{(2)\n_1}+\eta^{\n_3\n_1}p^{(3)\n_2}\biggr)p^{(1)\m_3}p^{(2)\m_1}p^{(3)\m_2} \Biggr\}.
\eeq 
We realized immediately that it could be identified with ${\cal A}_{hhh,C}^{(1,3)} + {\cal A}_{hhh,C}^{(3,1)}$ of closed string.  
Further, simple algebraic calculations yielded
\beq
{\cal A}_{hhh,O}^{(0,6)} &=&\frac{\kappa}{2}
\int \prod_{i=1}^3 dp^{(r)} \d\left(\sum_{r=1}^3 p^{(r)} \right)  h_{\m_{1}\n_{1}}(p^{(1)})h_{\m_{2}\n_{2}}(p^{(2)})h_{\m_{3}\n_{3}}(p^{(3)})\nn\\
&& p^{(1)\m_3} p^{(2)\m_1} p^{(3)\m_2}p^{(1)\n_3} p^{(2)\n_1} p^{(3)\n_2}. 
\eeq 
We observed that ${\cal A}_{hhh,O}^{(0,6)}$ agreed exactly with $\frac{1}{2}{\cal A}_{hhh,C}^{(3,3)}$ of closed string theory.
Combining them together, we find the difference between the cubic interactions of the massive symmetric rank-two tensor in open string theory and those of graviton in closed string theory:
\beq
{\cal A}_{hhh,O} (h) - {\cal A}_{hhh,C} (h) &=& {\cal A}_{hhh,O}^{(3,0)}(h) - \half {\cal A}_{hhh,O}^{(0,6)} + 2\kappa
\int \prod_{i=1}^3 dp^{(r)} \d\left(\sum_{r=1}^3 p^{(r)} \right)  h_{\m_{1}\n_{1}}(1)h_{\m_{2}\n_{2}}(2)h_{\m_{3}\n_{3}}(3) \nn\\
&& \Biggl\{\eta^{\m_1\m_2}p^{(1)\m_3}\eta^{\n_2\n_3}p^{(2)\n_1}+ \eta^{\m_2\m_3}p^{(2)\m_1}\eta^{\n_3\n_1}p^{(3)\n_2}
+\eta^{\m_3\m_1}p^{(3)\m_2}\eta^{\n_2\n_3}p^{(1)\n_3}\Biggr\}.
\eeq
As $\ap \rightarrow \infty$, the symmetric rank-two tensor becomes massless and 
the cubic interactions in the massless limit behave as follows:
\beq
{\cal A}_{hhh,O}^{(3,0)} \sim O((\ap)^{-3/2}), ~~{\cal A}_{hhh,O}^{(2,2)} \sim O((\ap)^{-1/2}), ~~
{\cal A}_{hhh,O}^{(1,4)} \sim O((\ap)^{1/2}), ~~{\cal A}_{hhh,O}^{(0,6)} \sim O((\ap)^{3/2}).
\eeq 
This observation leads to the conclusion that in the massless limit, (or equivalently high energy limit), the $\ap$-correction terms,
containing higher derivatives dominate in both string theories: ${\cal A}_{hhh,O}^{(0,6)}$ of open string theory and  
${\cal A}_{hhh,C}^{(3,3)}$ of closed string theory which in fact are identical.

\section{Discussions and Conclusions}

We evaluated three particle-scattering amplitudes of spin-two particles in both closed and open string theories 
using Polyakov's string path integral in the proper-time gauge. 
Cubic couplings of spin-two particles were directly obtained from the three particle-scattering 
amplitudes. As for the three-graviton interactions, including $\ap$-corrections, we confirmed a previous result \cite{GreenSW1987}, 
which was completely factorized in accordance with the KLT relations \cite{Kawai86}. 
Then, we calculated three-particle scattering amplitudes of the
massive symmetric rank-two tensor, carrying spin two, in open string theory to derive the cubic couplings of massive spin-two particles. 
We compared the cubic interactions of two spin-two particles and observed that the cubic coupling of the massive symmetric rank-two 
tensor differs from that of the graviton. In the high-energy limit, where $\ap \rightarrow \infty$,
the massive symmetric rank-two tensor becomes massless, the $\ap$-correction term with six derivatives dominates. 
As a result, classical cubic actions of both theories become identical. 

An immediate extension of this work may be to calculate four-particle scattering amplitudes of
symmetric rank-two tensors and compare them with the four-graviton scattering amplitudes, which have been 
studied in a recent work \cite{TLee2019PLB}. It would also be interesting to further extend this work to higher-spin particles in string theory \cite{Sagnotti2011,Metsaev2012,Joung2014}
to examine their high-energy limits \cite{Gross87,Gross88} as suggested elsewhere. The Fock space representations of three-string scattering amplitudes given in Eq. (\ref{3string}) and Eq. (\ref{open3}) would be useful to study higher-spin particles in general.

\vskip 1cm

\begin{acknowledgments}

TL was supported by Basic Science Research Program through the National Research Foundation of Korea(NRF) funded by the Ministry of Education (2017R1D1A1A02017805)
and also by the 2018 Research Grant (PoINT) from Kangwon National University.
\end{acknowledgments}


%

%





\end{document}